\documentclass[11pt,a4paper]{article}

\usepackage[T1]{fontenc}
\usepackage[utf8]{inputenc}
\usepackage{lmodern}
\usepackage[margin=1in]{geometry}
\usepackage{microtype}
\usepackage{amsmath,amssymb,amsthm,mathtools}
\usepackage{physics}
\usepackage{tikz}
\usepackage{booktabs}
\usepackage{enumitem}
\usepackage[hidelinks]{hyperref}
\usepackage[nameinlink,capitalize]{cleveref}
\usepackage{xcolor}
\usepackage{graphicx}
\usepackage{listings}
\usepackage{framed}

\lstdefinelanguage{Mathematica}{
  morekeywords={BeginPackage,EndPackage,Module,Block,With,If,For,While,Return,Print,Map,Select,ReplaceAll,Association,AssociationQ,KeyExistsQ,Import,Export,Options,OptionValue,Table,Range,MatrixForm,Normal,Series,Expand,Simplify,Coefficient,CoefficientArrays},
  sensitive=true,
  morecomment=[l]{(*},
  morecomment=[s]{(*}{*)},
  morestring=[b]"
}
\theoremstyle{definition}

\usepackage[many]{tcolorbox}
\tcbset{colback=gray!3,colframe=gray!50!black,boxrule=0.2pt,arc=1mm}

\newcommand{\PP}{\mathbb{P}}

\newcommand{\Pic}{\operatorname{Pic}}
\newcommand{\Sym}{\operatorname{Sym}}

\usepackage{xcolor}

\newlength{\myboxbeforeskip} 
\newlength{\myboxafterskip}  
\newlength{\myboxframesep}   
\newlength{\myboxframerule}  
\newcommand\myboxbg{yellow!20}

\newsavebox{\myboxbox}

\newcommand{\mybox}[1]{%
  \refstepcounter{equation}%
  \par\vspace{\myboxbeforeskip}\noindent
  \sbox{\myboxbox}{%
    \begingroup
      \setlength{\fboxsep}{\myboxframesep}%
      \setlength{\fboxrule}{\myboxframerule}%
      \fcolorbox{black}{\myboxbg}{%
        \ttfamily
        \begin{tabular}{@{}c@{}}%
          \strut #1%
        \end{tabular}%
      }%
    \endgroup
  }%
  \usebox{\myboxbox}%
  \hfill(\theequation)%
  \par\vspace{\myboxafterskip}%
}
 \numberwithin{equation}{section}

\usepackage[T1]{fontenc}
\usepackage[utf8]{inputenc}
\usepackage[svgnames, table]{xcolor}
\colorlet{framecolor}{Gold}
\colorlet{backcolor}{Orange}

\usepackage{eqparbox}
\usepackage{makecell}
\usepackage{bigstrut}

\title{\vspace{-0.5em}\textbf{CIPro Package: \\ Complete Intersections in Products of Projective Spaces and Line Bundles}
}
\date{}

\begin{document}
\phantom{}
\vspace{-.2cm}

\begin{center}
{\LARGE {\bf {\Large\bfseries CIPro Package: Complete Intersections\\[4pt] in Products of Projective Spaces and Line Bundles}}}\\[24pt]
{\bf{Lara B.~Anderson$^{a,}$\footnote{lara.anderson@vt.edu},
Andrei Constantin$^{b, c,}$\footnote{andrei.constantin@physics.ox.ac.uk},\\[4pt] James Gray$^{a, }$\footnote{jamesgray@vt.edu}, Yang-Hui He$^{d,}$\footnote{yh@lims.uk}, Seung-Joo Lee$^{e,}$\footnote{seungjoolee@yonsei.ac.kr}, Andre Lukas$^{c,}$\footnote{lukas@physics.ox.ac.uk}}}
\bigskip\\[0pt]
\vspace{0.33cm}
${}^a$ {\it Physics Department, Robeson Hall, Virginia Tech,
Blacksburg, VA 24061, U.S.A.}
\vspace{8pt}

${}^b$ {\it 
School of Mathematics, University of Birmingham, Watson Building,\\ Edgbaston, Birmingham B15 2TT, United Kingdom
}
\vspace{8pt}

${}^c$ {\it 
Rudolf Peierls Centre for Theoretical Physics, University of Oxford\\
Parks Road, Oxford OX1 3PU, United Kingdom
}
\vspace{8pt}

${}^d$ {\it 
London Institute for Mathematical Sciences, Royal Institution,\\ London, W1S 4BS, UK
}
\vspace{8pt}

${}^e$ {\it 
Department of Physics, Yonsei University, Seoul 03722, Republic of Korea
}
\end{center}
\vspace{2.4cm}

\begin{abstract}
\noindent
\texttt{CIPro} is a Mathematica package for constructing and analyzing complete intersections in products of projective spaces and line bundles over such varieties. It computes properties of complete intersections, such as Chern classes, Hilbert series, GV invariants and symmetries, as well as properties of line bundles on complete intersections, including their cohomology groups. The package also consolidates a number of data sets available in the literature into a single system, including the lists of complete intersection Calabi--Yau three- and four-folds. This short tutorial introduces the package, provides a brief discussion of some of the mathematical background underlying its computations, and gives a series of examples to illustrate its use. These tools are of utility for many computations in string compactifications, especially for Calabi--Yau geometries appearing in Heterotic and Type II constructions. Many tools apply beyond the Calabi--Yau context, including for example, almost Fano bases in F-theory.

\end{abstract}
\vfill

\pagebreak
\tableofcontents

\newpage
\section{Overview, installation and system basics}
\label{sec:intro}

This paper introduces \texttt{CIPro}, a \texttt{Mathematica} package designed for the systematic study of complete intersection (CIs) varieties in products of projective spaces and of line bundles over those spaces. 
\texttt{CIPro} implements a range of routines for the study of geometric and topological quantities, including intersection numbers, Chern and Todd classes, canonical classes, Euler characteristics and Hilbert series. It further contains tools for constructing defining equations of given multi-degree, checking smoothness and singularity properties, and exploring automorphisms of the underlying varieties.

A distinguishing feature of the package is its integrated approach to line bundle cohomology. Using the Koszul spectral sequence in conjunction with Bott--Borel--Weil computations on the ambient space, \texttt{CIPro} computes the cohomology of (direct sums of) line bundles. These components form a coherent computational environment with applications in both mathematics and theoretical physics.

The package comes with a number of data sets which include the set of 7890 complete intersection Calabi-Yau three-folds in products of projective spaces (CICYs) from Ref.~\cite{Candelas:1987kf} and the analogous set of Calabi-Yau four-fold CICYs from Refs.~\cite{Gray:2013mja,Gray:2014kda}. In addition, there is data for freely-acting symmetries on CICY three-folds, taken from the classification in Ref.~\cite{Braun:2010vc}, and data with line bundle sums which lead to physically promising particle physics models in the context of heterotic string compactifications, taken from Refs.~\cite{Anderson:2011ns,Anderson:2012yf,Anderson:2013xka, Constantin:2018xkj}.

The tools available in this package are of particular use in the study of string compactifications and related geometry. Applications include not just Calabi-Yau 3- and 4-fold compactifications using the datasets described above, but also more general Calabi-Yau constructions (including those described in Ref.~\cite{Anderson:2015iia}) and the study of (not necessarily toric) almost Fano bases for F-theory in 6- and 4-dimensions (see, for instance, Ref.~\cite{Taylor:2015isa}) or backgrounds for Type II orientifolds (see, for instance, Ref.~\cite{Carta:2020ohw}).

\begin{framed}
\vspace{-0.5cm}
\section*{Installation and System Basics}

{\bf The package can be downloaded from this \textcolor{blue}{\href{https://github.com/JamesGray-Physics/CIPro/releases/latest}{github repository}.}}

To install CIPro, execute the following command in \texttt{Mathematica}:
\mybox{$\begin{aligned}
  \text{In[]}&:= \$\text{UserBaseDirectory}
\end{aligned}$}
This returns a path, \texttt{userbasedirectory}, to a folder on your computer. Download and unzip the package and place the four resulting items (CIPro.m and the three folders CIProAux, CIProData and CIProExamples) in the sub-directory \texttt{userbasedirectory/Applications/} of the  UserBaseDirectory. This concludes the installation.

To load the package execute the following command in a Mathematica notebook.
\mybox{$\begin{aligned}
  \text{In[]}&:= <<\text{CIPro.m}
\end{aligned}$}
The package is now loaded and ready to use.

\vspace{0.2cm}

The package primarily uses two modules, $\texttt{CIProp}$ and $\texttt{CIMod}$, to compute properties of complete intersection varieties and bundles over them. This document contains a tutorial of some of the basic functionality, but help is also available from within Mathematica. Executing the following commands
\mybox{$\begin{aligned}\label{help1}
 \text{In[]}&:=\text{CIPro[]}\\
  \text{In[]}&:= \text{CIProp[]} \\ 
  \text{In[]}&:= \text{CIMod[]} 
  \end{aligned}$}
will provide basic information and a list of quantities that the package can compute in terms of ``keywords.'' Longer versions of this help can be obtained by providing the argument \texttt{"Long"} to any of the commands in \eqref{help1}. The command
\mybox{$\begin{aligned}\label{help}
 \text{In[]}&:=\text{CIPro[\text{``Manual"}]}
\end{aligned}$}
brings up the present paper as a package manual. Information about a specific keyword \texttt{key} can be obtained by executing one of the commands
\mybox{$\begin{aligned}
  \text{In[]}&:=\text{CIProp[key]}\\
  \text{In[]}&:= \text{CIProp[\text{key},\text{``Long"}]}  \\
   \text{In[]}&:=\text{CIMod[key]}\\
  \text{In[]}&:= \text{CIMod[\text{key},\text{``Long"}]}  \end{aligned}$}
for a short or long version. The long version includes information on the mathematical background for  the property represented by \texttt{key} and a simple, executable, example.

\vspace{0.2cm}

We recommend looking through this tutorial as a first introduction to CIPro, as it gives a more cohesive overview to the structure of the package and an easier entry to computing the quantities that we anticipate being of interest to most users. 

\vspace{0.2cm}

Users making use of this package in published work are kindly requested to cite this manual.

\end{framed}

We note that a number of excellent and complementary software packages are available for studying various aspects of Calabi--Yau and related geometries. These include, but are not limited to, \cite{m2,singular,cohomcalg,pycicy,cytools,sagemath,grdb}.

\vspace{0.6cm}

In the following three chapters we will be presenting the main features of the package in tutorial style. In Section 2 we discuss how to set up complete intersections as Mathematica associations and how to compute their properties, including characteristic classes, defining equations, Hilbert series, intersection forms and information on singularities. Section 3 explains how to add line bundles over a complete intersection to the set-up and compute their properties, including characteristic classes and cohomology. Finally, in Section 4 we explain how to introduce complete intersections with (finite) automorphism groups. A full list of computable properties and available datasets is provided in the Appendices. For more comprehensive information about the mathematical background we recommend Refs.~\cite{hubsch,schenck,cox,gh,h}. The package uses an algorithm for computing line bundle cohomology on CIs which is based on the Bott--Borel--Weil formalism for line bundle cohomology on projective spaces. We have included a brief account of the algorithm in Appendix~\ref{app:lbc}.

\subsection{Key Features}
We summarize some of the key features of \texttt{CIPro} and highlight potential applications in both mathematics and theoretical physics.
\begin{description}
    \item[Basic geometric properties:] The package can determine basic geometric properties of a CI, including whether it is Calabi--Yau and whether it is smooth. In the latter case, information about the singular locus can also be obtained.

    \item[Chern Classes: ] The package computes the Chern polynomial of the CI in terms of the K\"ahler forms of the ambient product of projective spaces. Related invariants, such as the Euler characteristic, Todd classes, and triple intersection numbers, can also be determined.
    
    \item[Hilbert Series: ] The multigraded Hilbert series is a fundamental object associated with the graded coordinate ring of an embedded variety. It also plays a central role in counting gauge-invariant operators in supersymmetric gauge theories \cite{Benvenuti:2006qr}.

    \item[GV and GW Invariants: ]The package allows the computation of Gopakumar--Vafa invariants for CICYs by integrating code developed by Albrecht Klemm based on Refs.~\cite{Hosono:1993qy,Hosono:1994ax}. The corresponding Gromov--Witten invariants can then be obtained from the Gopakumar--Vafa invariants via standard techniques. These invariants play a central role in modern enumerative geometry \cite{Hori:2003ic}.

    \item[Line Bundle Topology and Cohomology: ]
    Line bundles, as well as direct sums thereof, can be defined on the CI. Topological and cohomological quantities, including Chern classes, bundle cohomology, and the associated Koszul complex, can be computed for arbitrary user-specified configurations.

    \item[Symmetries:] The package supports basic finite-group computations, including group actions on CIs and the construction of CIs invariant under a prescribed finite symmetry. 
\end{description}

\section{Complete intersections and their properties}
\label{sec:definitions}

Let $\mathcal A = \PP^{n_1} \times \PP^{n_2} \times \cdots \times \PP^{n_m}$ be a product of complex projective spaces, called the \emph{ambient space}. In order to define a complete intersection we should consider line bundles and their global sections on $\mathcal{A}$. Denote by $\pi_i : \mathcal A \to \PP^{n_i}$ the canonical projections, and by
\begin{equation}\label{lbamb}
\mathcal{O}_{\mathcal A}(\mathbf{k}) := 
  \bigotimes_{i=1}^m \pi_i^*\mathcal{O}_{\PP^{n_i}}(k^i),
  \qquad \mathbf{k} = (k^1,\dots,k^m)^T \in \mathbb{Z}^m,
\end{equation}
the standard multi-graded line bundle on $\mathcal  A$.  The  global sections, $\Gamma(\mathcal{A},\mathcal{O}_{\mathcal A}(\mathbf{k}))$ consist of  multi-homogeneous polynomial of multi-degree ${\bf k}$. More precisely, this means such polynomials are homogeneous of degree $k^i$ in the homogeneous coordinates of the $i^{\rm th}$ factor $\PP^{n_i}$.

\subsection{Definition and configuration matrix} \label{defsec}
In order to define a complete intersection of co-dimension $r$ in $\mathcal{A}$ we should select $r$ multi-degrees ${\bf q}_a\in\mathbb{Z}^m_{\geq 0}$. Choosing corresponding global sections $s_a\in\Gamma(\mathcal{A},{\cal O}_{\mathcal{A}}({\bf q}_a))$, a \emph{complete intersection variety with multi-degrees} $({\bf q}_1,\ldots ,{\bf q}_r)$ is then defined as the common zero locus of these sections, that is,
\begin{equation}\label{Xdef}
X = V(s_1, \dots, s_r)
  := \{ x \in \mathcal A \mid s_1(x) = \cdots = s_r(x) = 0 \}\; .
\end{equation}
It is convenient to record the multi-degrees in a single \emph{configuration matrix},
\begin{equation}
 (q_a^{i}) = \left[
\begin{array}{c|cccc}
\PP^{n_1} & q_1^{1} & q_2^{1} & \cdots & q_r^{1}\\
\PP^{n_2} & q_1^{2} & q_2^{2} & \cdots & q_r^{2}\\
\vdots    & \vdots    & \vdots    & \ddots & \vdots   \\
\PP^{n_m} & q_1^{m} & q_2^{m} & \cdots & q_r^{m}
\end{array}
\right]\; ,
\end{equation}
where the $i^{\rm th}$ row contains the information related to the ambient space factor $\PP^{n_i}$  and whose $a^{\rm th}$ column contains the multi-degree ${\bf q}_a$ of the section $s_a\in H^0\!\big(\mathcal{A},\mathcal{O}_{\mathcal{A}}(\mathbf{q}_a)\big)$. It is also useful to define the line bundle sum $\mathcal{N}=\bigoplus_{a=1}^r{\mathcal O}_{\mathcal{A}}({\bf q}_a)$ whose restriction $N=\mathcal{N}|_X$ is the normal bundle of the CI.

\subsubsection*{Example:}

Suppose we would like to set up the following description of the $dP_2$ surface within the package.
\begin{eqnarray} \label{dp21}
dP_2 \in \left[\begin{array}{c|cc} \mathbb{P}^2& 1&1 \\ \mathbb{P}^1&1&0 \\ \mathbb{P}^1&0&1\end{array} \right]
\end{eqnarray}
This can be input by hand as follows.

\mybox{$\begin{aligned}
  \text{In[]}&:= \text{dp2}=<|\text{``DimPs''}\to\{2,1,1\},\text{``Conf''}\to\{\{1,1\},\{1,0\},\{0,1\}\}|> ;
\end{aligned}$} \label{dp22}

This object is a Mathematica association, the basic data structure used in the package. As the package computes various properties of a given variety it will add them to the association describing that geometry. The association contains a set of entries of the form $\texttt{``Keyword''} \to $ value. Here, $\texttt{``DimPs''}$ is the keyword recording the dimensions of the ambient projective space factors as a Mathematica list $\{n_1,\ldots ,n_m\}$. The keyword for the configuration matrix itself is $\texttt{``Conf''}$ which provides the $m\times r$ configuration matrix $(q_a^i)$ as a Mathematica matrix. It is important that the rows of this matrix are provided in the same order in which the projective space factors are given in $\texttt{``DimPs''}$.

Alternatively, one can load a complete intersection from one of the data sets provided with the package. Here is an example of a CICY threefold, taken from the dataset named $\texttt{"Cicy3"}$, originally compiled in Ref.~\cite{Candelas:1987kf}.
\mybox{$\begin{aligned} \text{In[]}&:=  \text{CY} = \text{Data}[\text{``Cicy3''}, 7821] \\
\text{Out[]}&:=<|\text{``Num''}\to 7821, \text{``Conf''}\to \{\{1,1,1\},\{2,2,1\}\},\text{``Euler''}\to -112, \\ &\;\;\;\;\;\;\; \;\;\;\text{``H11''}\to 2,\text{``H21''}\to 58,\text{``C''}\to\{\{0,0\},\{36,50\},-112\} \, , \, \ldots|>\end{aligned}$}\label{cybox1}
This contains much more than just the defining data of the CICY, which we will not concern ourselves with here. The user interested in what these other keywords describe can execute the following command
\mybox{
\text{In[]}:=\text{Data}[\text{``Cicy3''}]
}
which provides general help and an explanation of the keywords for  the dataset named in the argument. Note, if the keyword $\texttt{``DimPs''}$ is not provided in an association, as in the example in (\ref{cybox1}), the dimensions of the ambient space projective factors are determined from the configuration matrix in $\texttt{``Conf''}$ by the Calabi-Yau condition, that is, $n_i=\sum_{a=1}^r q_a^i-1$.

\subsection{Smoothness}

Choose homogeneous coordinates
\begin{equation}
[x_0^{(i)}:\cdots:x_{n_i}^{(i)}]
\quad \text{on each factor } \PP^{n_i},
\end{equation}
so that $s_a$ may be represented by a multi-homogeneous polynomial in these coordinates for each $a$.  
At a point $x\in X$, consider the differentials
\begin{equation}
ds_a(x) \in T^*_x \mathcal{A}, \qquad a = 1,\dots,r .
\end{equation}
The variety $X$ is said to be \emph{smooth} if and only if the differentials
$\{ds_a(x)\}_{a=1}^r$ are linearly independent for all $x\in X$; equivalently,
the Jacobian matrix of partial derivatives of $(s_1,\dots,s_r)$ has rank $r$ everywhere on $X$:
\begin{equation}
{\rm rank}\left(\frac{\partial s_a}{\partial x_j^{(i)}}\right)_
{\substack{a=1,\ldots,r\\
i=1,\ldots,m,\;j=0,\ldots,n_i}}=r
\qquad\text{for all }x\in X.
\end{equation}

If this condition holds, then $X$ is a smooth projective variety of dimension
\begin{equation}
\dim X = \sum_{i=1}^m n_i - r .
\end{equation}

\subsubsection*{Example:}

Consider the $dP_2$ example given in (\ref{dp21}) and (\ref{dp22}). We might wish to know if a given choice of defining relations leads to a smooth variety. Say we label the coordinates of the $\mathbb{P}^2$ and two $\mathbb{P}^1$ factors in (\ref{dp21}) as $x=(x_0,x_1,x_2)$, $y=(y_0,y_1)$ and $z=(z_0,z_1)$ and that we wish to check smoothness of the variety defined as the common zero locus of
\begin{eqnarray}
s_1 = x_0 y_0 \;\;\;, \;\;\; s_2 = x_0 z_1 \; .
\end{eqnarray}
To do so we can specify coordinate names and defining relations in the association describing the complete intersection as follows.
\mybox{$\begin{aligned} 
\label{dP2def}
  \text{In[]}&:= \text{dp2}=<|\text{``DimPs''}\to\{2,1,1\},\text{``Conf''}\to\{\{1,1\},\{1,0\},\{0,1\}\}, \\
  &\;\;\quad\quad \quad \quad\quad \text{``AmbCoords''} \to \{\{x_0,x_1,x_2\}, \{y_0,y_1\}, \{z_0,z_1\}\},\\& \;\;\quad\quad \quad \quad\quad \text{``DefEqs''}  \to \{x_0 y_0, x_0 z_1\}|> ;
\end{aligned}$}\label{dp23}
To compute a property of a CI one calls the module $\texttt{CIProp}$, with the CI association and an appropriate keyword \texttt{key} as input. The output is another CI association which contains  the content of the input association plus the additional information \texttt{key}$\to$\texttt{value} for the computed quantity. The appropriate keyword for the task at hand is $\texttt{``Singular''}$ and singularity of the $dP_2$ defined in \eqref{dP2def} is checked as follows.
\mybox{$\begin{aligned} 
  \text{In[]}&:=\text{CIProp}[\text{dp2},\text{``Singular''}]; \\
  \text{Out[]}&:=<|\text{``DimPs''}\to\{2,1,1\},\text{``Conf''}\to\{\{1,1\},\{1,0\},\{0,1\}\}, \\
  & \;\;\; \quad\quad\text{``AmbCoords''} \to \{\{x_0,x_1,x_2\}, \{y_0,y_1\}, \{z_0,z_1\}\},\text{``DefEqs''}  \to \{x_0 y_0, x_0 z_1\} \\
& \;\;\; \quad \quad\text{``MonList''} \to \{\{x_0 y_0, 
   x_0 y_1, x_1y_0, 
   x_1 y_1, x_2 y_0, 
   x_2 y_1\}, \ldots\}, \\&\;\; \quad \quad \text{``Singular''} \to \text{True} |>
\end{aligned}$}\label{dp24}
We now see that the $\texttt{``Singular''}$ keyword has indeed been set to $\texttt{``True''}$ in the association.

As another example, we can check whether a generic CY three-fold in the family  specified in (\ref{cybox1}) is smooth. If  defining relations are not explicitly provided, the package will generate random, generic examples for a standard choice of coordinate names and add them to the association before performing the smoothness computation.
\mybox{$\begin{aligned} 
  \text{In[]}&:=\text{CIProp}[\text{CY},\text{``Singular''}][\text{``Singular''}]; \\
  \text{Out[]}&:= \text{False} 
\end{aligned}$}
Note that, because the defining relations are quite long in this case, we have chosen to only output the keyword of interest by adding $\texttt{[``Singular'']}$ to the command. The above is our first example of another general feature of the package. If the package is not given information which is required for the computation of a requested keyword, such as the names of the coordinates and the defining section in the above example, the required dependencies are calculated (or set to default choices) first and the information is added to the association.

\subsection{Chern Polynomials and other topological properties}

The tangent bundle $TX$ of the CI fits into the short exact sequence
\[
 0\rightarrow TX\rightarrow T{\cal A}|_X\rightarrow{\cal E}|_X\rightarrow 0\quad\mbox{where}\quad  {\cal E}=\bigoplus_{a=1}^r{\cal O}_{\cal A}({\bf q}_a)\; .
\]
Using $c(T{\cal A})=\wedge_{i=1}^m(1+J_i)^{n_i+1}$ as well as $c({\cal E})=\wedge_{a=1}^r(1+q_a^jJ_j)$, where $(J_1,\ldots ,J_m)$ is the standard set of forms obtained from the K\"ahler forms of the ambient space projective factors (normalised as $\int_{\mathbb{P}^{n_i}}J_i^{n_i}=1$), the above sequence implies that
\[
 c(TX)=\frac{c(T{\cal A})}{c(T{\cal E})}=\frac{\wedge_{i=1}^m(1+J_i)^{n_i+1}}{\wedge_{a=1}^r(1+q_a^jJ_j)}=1+c_1^iJ_i+c_2^{ij}J_i\wedge J_j+\cdots\; .
\]
Expanding this expression gives the Chern class $c(TX)$ as a polynomial in the forms $J_i$. 

\subsubsection*{Example:}

The computation of the Chern Polynomial is as straightforward as any other property the package accommodates. One simply calls $\texttt{CIProp}$ with the desired keyword. Here is a calculation of the total Chern class, represented by the keyword \texttt{"CPoly"}, for the  $dP_2$ example defined in (\ref{dp21}) and (\ref{dp22}).
\mybox{$\begin{aligned} 
  \text{In[]}&:=\text{CIProp}[\text{dp2},\text{``CPoly''}]\\
  \text{Out[]}&:= <|\text{``DimPs''}\to\{2,1,1\},\text{``Conf''}\to\{\{1,1\},\{1,0\},\{0,1\}\}, \text{``AmbDim''}\to 4 \\
  & \;\;\;\quad\quad \text{``CIDim''}\to 2, \text{``Mu''}\to (J_1+J_2)(J_1+J_3), \\& \;\;\;\quad\quad \text{``CPoly''}\to 1+J_1+J_2+ 2J_1 J_2+J_3+2 J_1 J_3+J_2 J_3|>
\end{aligned}$}

Other properties of complete intersections can be computed in an analogous manner and, as before, one can simply print out only the requested quantity if desired. The following, for example, computes the Hilbert Series of the CY manifold defined in (\ref{cybox1}).
\mybox{$\begin{aligned} 
  \text{In[]}&:=\text{CIProp}[\text{CY},\text{``HilbertSeries''}][\text{``HilbertSeries''}]\\
  \text{Out[]}&:= \frac{(1-t_1 t_2)(1-t_1 t_2^2)^2}{(1-t_1)^3(1-t_2)^5}
\end{aligned}$}
In terms of properties of complete intersection varieties themselves, the package can compute Chern characters and characteristics, Euler numbers, Hilbert series, intersection numbers and forms, canonical classes, Todd polynomials and Gopakumar--Vafa invariants (with the last of these using code by Albrecht Klemm based on Refs.~\cite{Hosono:1993qy,Hosono:1994ax}). Appendix~\ref{app:modules} contains a complete list of the relevant keywords.

 \subsection{Redundancy and normal forms of configurations}
 
 It is possible for different configuration matrices to describe the same algebraic variety. One simple way in which this can happen is if two such matrices are related by row and column permutations. \texttt{CIPro} has a command to put configuration matrices in a standard form so that such redundancies can be detected.
 
 Consider for example the configuration matrix describing $dP_2$ given in (\ref{dp21}) and (\ref{dp22}).
 \mybox{$\begin{aligned} 
  \text{In[]}&:=\text{dp2}[\text{``Conf''}]//\text{MM} \\ 
  \text{Out[]}&:= \left( \begin{array}{cc} 1&1\\1&0\\0&1\end{array} \right)
\end{aligned}$}
Here we have used the package's display command \texttt{MM} to put the configuration matrix in an easily readable form. This configuration matrix can be put in the package's normal form as follows.
 \mybox{$\begin{aligned} 
  \text{In[]}&:=\text{CIMod}[\text{dp2}, \text{``CINormal"}] // \text{MM} \\
  \text{Out[]}&:=<| \text{``Conf"}\to \left(
\begin{array}{cc}
 1 & 0 \\
 0 & 1 \\
 1 & 1 \\
\end{array}
\right),\text{``DimPs"}\to \{1,1,2\},\\& \quad \quad \;\text{``RowPerm"}\to \left(
\begin{array}{ccc}
 0 & 1 & 0 \\
 0 & 0 & 1 \\
 1 & 0 & 0 \\
\end{array}
\right),\text{``ColPerm"}\to \left(
\begin{array}{cc}
 1 & 0 \\
 0 & 1 \\
\end{array}
\right)|>
\end{aligned}$} \label{nf1}
Note we have used $\texttt{CIMod}$, rather than $\texttt{CIProp}$. The module $\texttt{CIProp}$, as discussed earlier in this section, is designed to add data to a CI  association. By contrast, the module $\texttt{CIMod}$ is the package's general purpose module for computing a new CI association or a different output type from a given CI association. Because the output in the case at hand is not simply an addition to the input association, but rather a new structure altogether, the $\texttt{CIMod}$ command has to be used. The  keyword \texttt{"Conf"} in the output assocation now points to the normal form of the configuration matrix. The keyword \texttt{"DimPs"} refers to the ambient space dimensions which have been re-ordered accordingly and the last two entries provide the row and column permutation required to produce the normal form of the configuration from the input one. The normal form of any configuration matrix related by row and column permutations will of course be the same. 

This means the normal form functionality allows the package to decide if two configuration are equivalent, in the sense that its configuration matrices and ambient space dimensions are  related by row and column permutations. For example, the following two configuration matrices for $dP_2$ are recognized as equivalent:
 \mybox{$\begin{aligned} 
  \text{In[]}:=&\text{alsodp2}=<|\text{``DimPs''}\to\{1,1,2\},\text{``Conf''}\to\{\{0,1\},\{1,0\},\{1,1\}\}|> ;\\
  &\text{CIMod}[\{\text{dp2},\text{alsodp2}\},\text{``CIEquiv"}] \\ 
  \text{Out[]}:=& \text{True}
\end{aligned}$}

\section{Line bundles on complete intersections}

Frequently, one is interested not only in the geometry of an algebraic variety itself, but also in other structures defined over that variety. For example, vector bundles over varieties are used extensively in a plethora of settings. In this section, we describe how \texttt{CIPro} can be used to compute a large array of properties of vector bundles, using line bundles as a basic building block.

\subsection{Line bundles}

Every line bundle on $\mathcal{A}$ of the form $\mathcal{O}_{\mathcal{A}}(\mathbf{k})$ restricts to a line bundle on $X$:
\begin{equation}
\mathcal{O}_X(\mathbf{k}) := \mathcal{O}_{\mathcal{A}}(\mathbf{k})|_X .
\end{equation}
These are called \emph{multi-degree line bundles} on $X$.
Their first Chern class is
\begin{equation}
c_1\big(\mathcal{O}_X(\mathbf{k})\big)
  = \sum_{i=1}^m k^i\, J_i|_X \in H^2(X,\mathbb{Z}) .
\end{equation}
The group of all such line bundles forms a free abelian group
$\Pic(\mathcal{A}) \cong \mathbb{Z}^m$, and its image in $\Pic(X)$ has rank
$h^{1,1}(X)\le m$.  
When the restriction map $\Pic({\mathcal{A}})\to\Pic(X)$ is surjective,
we say that the complete intersection $X$ is \emph{favourable}.

Tensor operations are componentwise:
\begin{equation}
\mathcal{O}_X(\mathbf{k}) \otimes \mathcal{O}_X(\mathbf{l})
  = \mathcal{O}_X(\mathbf{k}+\mathbf{l}),
  \qquad
  \mathcal{O}_X(\mathbf{k})^\vee
  = \mathcal{O}_X(-\mathbf{k}),
  \qquad
  \Sym^p \mathcal{O}_X(\mathbf{k})
  = \mathcal{O}_X(p\mathbf{k}) .
\end{equation}

\subsubsection*{Example}

Say one wanted to work with the  complete intersection Calabi-Yau given by,
\begin{eqnarray} \label{lbseg1}
   X= \left[ \begin{array}{c|ccc}\mathbb{P}^2 & 2&1\\ \mathbb{P}^3 &1&3 \end{array} \right] \;,
\end{eqnarray}
with the following line bundle sum over it.
\begin{eqnarray} \label{lbseg2}
 {\cal O}_X(1,-1) \oplus {\cal O}_X (-5,2)
\end{eqnarray}
This structure can be described in the package by the following association.
 \mybox{$\begin{aligned} 
  \text{In[]}&:=\text{CYandLBS} = <|\text{``DimPs''} \to \{2, 3\}, \text{``Conf''} \to \{\{2, 1\},\{1,3\}\},\\& 
  \quad\quad\quad\quad\quad\quad\quad\quad \text{``LBS''} \to \{\{1,-5\},\{-1,2\}\}|>;
\end{aligned}$}\label{lbseg}

\subsection{Cohomology of line bundles}

For a line bundle $\mathcal{O}_X(\mathbf{k})$ on a smooth complete intersection $X$, one is often interested in the cohomology groups
\begin{equation}
H^q(X, \mathcal{O}_X(\mathbf{k})), \qquad q = 0, \dots, \dim X .
\end{equation}
These can be computed via the \emph{Koszul resolution}
\begin{equation}
0 \longrightarrow \wedge^r E \otimes \mathcal{O}_{\mathcal{A}}(\mathbf{k})
  \longrightarrow \cdots
  \longrightarrow
  E \otimes \mathcal{O}_{\mathcal{A}}(\mathbf{k})
  \longrightarrow
  \mathcal{O}_{\mathcal{A}}(\mathbf{k})
  \longrightarrow
  \mathcal{O}_X(\mathbf{k})
  \longrightarrow 0 ,
\end{equation}
where $E = \bigoplus_{a=1}^r \mathcal{O}_{\mathcal{A}}(\mathbf{q}_a)$.
The associated spectral sequence expresses $H^\bullet(X,\mathcal{O}_X(\mathbf{k}))$
in terms of the ambient cohomology groups
$H^\bullet({\mathcal{A}}, \wedge^p E \otimes \mathcal{O}_{\mathcal{A}}(\mathbf{k}))$,
which can be evaluated using the Bott--Borel--Weil theorem on each projective factor
and the K\"unneth formula for the product ${\mathcal{A}}$. For more details see Appendix~\ref{app:lbc}.

\subsubsection*{Example}

If one wishes to compute the cohomology of the line bundle sum given in (\ref{lbseg2}) over the base manifold (\ref{lbseg1}), as described by the association (\ref{lbseg}), one simply executes the following command. 
 \mybox{$\begin{aligned} 
  \text{In[]}&:=\text{CIMod}[\text{CYandLBS}, \text{``CohLBS''}] \\
  \text{Out[]}&:=\{ \{0 \},\{ 3\},\{4 \},\{ 0\}\}
\end{aligned}$}\label{output1}
The output given is in the form $\{\{h^0 \},\{h^1\},\{h^2\},\{h^3\}\}$, where $h^i$ is the dimension of the $i$'th cohomology group of the line bundle sum. If no defining relations are specified, as was the case in the above, then a generic defining relation is generated and used by the package. If, however, a specific defining relation is desired, this can easily be input. For example, let us choose the following defining relations for the Calabi-Yau threefold (\ref{lbseg1}).
\begin{equation}\label{defrelex}
\begin{array}{rcl}
p_{2,1}&=& x_{1,0}^2 x_{2,0}+x_{1,1}^2 x_{2,1} + x_{1,2}^2 x_{2,2}+x_{1,0}x_{1,1} x_{2,3} \\[2mm] 
p_{1,3}&=& x_{1,0} x_{2,3}^3+x_{1,1}x_{2,2}^3+x_{1,0}x_{2,1}^3+x_{1,2}x_{2,0}^3 +x_{1,2} x_{2,0} x_{2,1} x_{2,3}
\end{array}
\end{equation}

We can input this and then recompute the cohomology as follows. 
 \mybox{$\begin{aligned} 
  \text{In[]}&:=\text{newCYandLBS}=\text{CIProp}\left[\text{CYandLBS},\text{``DefEqs"},\right.\\ & \text{``Eqs"}\to \left\{ x_{1,0}^2x_{2,0}+x_{1,1}^2 x_{2,1}+x_{1,2}^2 x_{2,2}+x_{1,0} x_{1,1} x_{2,3} , \right. \\ & \left. \left. x_{1,0} x_{2,3}^3+x_{1,1} x_{2,2}^3+x_{1,0} x_{2,1}^3+x_{1,2} x_{2,0}^3+x_{1,2} x_{2,0}x_{2,1} x_{2,3} \right\}, \right.\\ & \left.\text{``AmbCoords"}\to \left\{\left\{x_{1,0},x_{1,1},x_{1,2}\right\},\left\{x_{2,0},x_{2,1},x_{2,2},x_{2,3}\right\}\right\}\right]; \\ 
  \text{In[]}&:=\text{CIMod}[\text{newCYandLBS}, \text{``CohLBS"}] \\
  \text{Out[]}&:=\{ \{0 \},\{ 4\},\{5 \},\{ 0\}\}
\end{aligned}$}\label{output2}
The fact that the cohomologies found in outputs (\ref{output1}) and (\ref{output2}) are different is due to the fact that the cohomology of ${\cal O}_X(-5,2)$ jumps in dimension at the special locus in moduli space represented by the defining polynomials in Eq.~\eqref{defrelex}. That this is a smooth point in moduli space can of course easily be checked:
\mybox{$\begin{aligned} 
  \text{In[]}&:=\text{CIProp}[\text{newCYandLBS},\text{``Singular''}][\text{``Singular''}]; \\
  \text{Out[]}&:= \text{False} 
\end{aligned}$}

\subsection{Images, kernels and cokernels of maps between cohomology groups}

Given a short exact sequence of vector bundles over a manifold $X$,
\begin{eqnarray}
0 \to A \to B \to C \to 0
\end{eqnarray}
there is an associated long exact sequence in cohomology.
\begin{eqnarray} \label{lesgen}
 0 \to H^0(X,A) \to H^0(X,B) \to H^0(X, C) \to H^1(X, A) \to \ldots \to H^{\text{dim}X}(X, C) \to 0
\end{eqnarray}
In analyzing the implications of such sequences, we frequently wish to compute the images, kernels and cokernels of maps between cohomology groups.

\subsubsection*{Example}

Consider the bicubic Calabi-Yau threefold.
\begin{eqnarray}
X= \left[ \begin{array}{c|c} \mathbb{P}^2 & 3 \\ \mathbb{P}^2 & 3\end{array} \right]
\end{eqnarray}
Say one decided to define a non-abelian bundle $V_X$ on this variety by the following short exact monad sequence.
\begin{eqnarray} \label{elmonad}
0 \to V_X \to {\cal O}_X(1,0)^3 \oplus {\cal O}_X(0,1)^3 \to {\cal O}_X(3,3)\to 0
\end{eqnarray}
This sequence leads to a long exact sequence in cohomology of the form (\ref{lesgen}). It is easy to verify in this case that $h^1(X,{\cal O}_X(1,0)^3 \oplus {\cal O}_X(0,1)^3)=0$.
 \mybox{$\begin{aligned} 
  \text{In[]}&:=\text{LBSone} = <|\text{``Conf"} \to \{\{3\}, \{3\}\}, \text{``DimPs"} \to \{2, 2\}, \\ &
 \quad\quad\quad\quad\quad\quad \quad \text{``LBS"} \to \{\{1, 1, 1, 0, 0, 0\}, \{0, 0, 0, 1, 1, 1\}\}|>;  \\ 
 \text{In[]}&:=\text{CIMod}[\text{LBSone},\text{``CohLBS"}] \\
 \text{Out[]}&:= \{\{18\},\{0\},\{0\},\{0\}\}
\end{aligned}$}
Therefore, the cohomology group $H^1(X, V_X)$ is given by the following co-kernel.
\begin{eqnarray} \label{mrwantywant}
H^1(X,V_X) = \text{Coker}\left[ H^0(X,{\cal O}_X(1,0)^3 \oplus {\cal O}_X(0,1)^3) \to H^0(X,{\cal O}_X(3,3))\right]
\end{eqnarray}
The higher level functionality in $\texttt{CIPro}$ allows the computation of such co-kernels (and indeed kernels and images of maps).

The procedure to compute a quantity such as (\ref{mrwantywant}) is to first obtain a full description of the source and target cohomologies, and then to compute the desired property of the map between them. For the case at hand, we proceed as follows. First we define the line bundle sums whose cohomologies will form the source and target spaces and append our choice of defining relations to those associations.
 \mybox{$\begin{aligned} 
  \text{In[]}&:=\text{source1} = <|\text{``Conf"} \to \{\{3\}, \{3\}\}, \text{``DimPs"} \to \{2, 2\}, \\& \quad\quad\quad\quad\quad\quad \quad 
   \text{``LBS"} \to \{\{1, 1, 1, 0, 0, 0\}, \{0, 0, 0, 1, 1, 1\}\}|>; \\&
\quad\;\; \text{target1} = <|\text{``Conf"} \to \{\{3\}, \{3\}\}, \text{``DimPs"} \to \{2, 2\}, 
   \text{``LBS"} \to \{\{3\}, \{3\}\}|>;  \\
   \text{In[]}&:= \text{source2}=\text{CIProp}[\text{source1}, \text{``DefEqs"}]; \\
   \text{In[]}&:= \text{target2} = \text{CIProp}[\text{target1}, \text{``DefEqs"}, \text{``Eqs"} \to \text{source2}[\text{``DefEqs"}], 
\\& \quad\quad\quad\quad\quad\quad \quad  \text{``AmbCoords"} \to \text{source2}[\text{``AmbCoords"}]];
\end{aligned}$}
Here note that we have been careful to use the same defining relations in the associations $\texttt{source2}$ and $\texttt{target2}$ that we will now use to compute the two different cohomologies on $X$. Next, we compute the source and target cohomology groups.
 \mybox{$\begin{aligned} 
  \text{In[]}&:=\text{sourceCoh} = \text{CIMod}[\text{source2},\text{``CohLBS"},\text{``CohFormat"}\to \text{``Long"}][[1]];\\
  \text{In[]}&:=\text{targetCoh} = \text{CIMod}[\text{target2},\text{``CohLBS"},\text{``CohFormat"}\to \text{``Long"}][[1]];
\end{aligned}$}
Here the $\texttt{``CohFormat"}\to \texttt{``Long"}$ options tells the package to output a detailed description of the cohomology, and not just its dimension, as will be needed to perform the mapping. The $[[1]]$ entries at the end of each call just tells Mathematica to keep the first entry in the list of cohomologies, that is $H^0$. Finally, we tell the package to compute the desired cokernel.
 \mybox{$\begin{aligned} 
  \text{In[]}&:=\text{CIMod}[\{\text{sourceCoh},\text{targetCoh}\}, \text{``MapSpaceLBS"}, 
 \\& \quad\quad\quad\quad\quad\quad \quad \text{``Space"} \to \text{``Coker"}] [\text{``LBSCohDim"}] \\
 \text{Out[]}&:= 81
\end{aligned}$}
Here we have only printed out the desired result --  in this case the dimension of the co-kernel given by $\texttt{``LBSCohDim''}$. A computation such as this one involves a large amount of data and explicit cohomology descriptions being derived behind the scenes and suppressing such information can be highly desirable.

As can be seen above, we did not specify a monad map for (\ref{elmonad}) in this example. In such a situation, the package chooses a generic map automatically. If a specific map is desired, this can be input by adding the optional argument $\texttt{``Map''} \to \texttt{matrix}$ where $\texttt{matrix}$ is an appropriate map for the given monad (here a $1\times 6$ matrix with three bi-degree $(2,3)$ and three bi-degree $(3,2)$ entries). 

{\bf An important note:} in computing such maps while utilizing the $\texttt{``MapSpaceLBS''}$ and $\texttt{``MapLBS''}$ keywords, the package sets to zero any maps corresponding to line bundles with negative entries that are nevertheless effective. In such cases, the map used is not, therefore, generic.

\subsection{Other properties of line bundle sums}

\texttt{CIPro} can compute many other properties of line bundle sums over complete intersection varieties, in addition to those discussed above. One might, for example be interested in various topological invariants, such as the Chern polynomial.

For a single line bundle $L={\cal O}_X({\bf k})$ on a complete intersection $X\subset{\cal A}$ with dimension $d$, we have $c_1(L)=k^iJ_i$, where $J_i$ are the standard forms, and ${\rm ch}(L)=1+\sum_{q=1}^d\frac{1}{q!}c_1(L)^q$. For a line bundle sum $U={\cal O}_X({\bf k}_1)\oplus\cdots\oplus{\cal O}_X({\bf k}_s)$ this implies
\[
 {\rm ch}(U)=\sum_{a=1}^s{\rm ch}({\cal O}_X({\bf k}_a))=r+\sum_{a=1}^s\sum_{q=1}^d\frac{1}{q!}(k_a^iJ_i)^q=s+{\rm ch}_1^i(U)J_i+{\rm ch}_2^{ij}(U)J_i\wedge J_j+\cdots
\]
\subsubsection*{Example}

Here is an example of the computation of the Chern polynomial of a line bundle sum over a complete intersection variety.
 \mybox{$\begin{aligned} 
  \text{In[]}&:=\text{CIandLBS} = <|\text{``DimPs"} \to \{2, 3\}, \text{``Conf"} \to \{\{2 d, 1\}, \{1, 3\}\}, \\& \quad \quad \quad \quad \quad  \quad \quad\quad 
  \text{``LBS"} \to \{\{1, 1, -5 a\}, \{b, -1, 2\}\}|>; \\ 
  \text{In[]}&:=\text{CIProp}[\text{CIandLBS}, \text{``ChLBSPoly"}][\text{``ChLBSPoly"}] \\
  \text{Out[]}&:=-\frac{1}{6} 125 a^3 J_1^3+\frac{25}{2} a^2 J_1^2+25 a^2 J_2 J_1^2-10 a J_2^2 J_1-5 a J_1-10 a J_2 J_1+\frac{1}{6} b^3 J_2^3 \\&\quad \;\;+\frac{1}{2} b^2 J_2^2 J_1+\frac{1}{2} b^2 J_2^2+\frac{1}{2} b J_2 J_1^2 +b J_2 J_1+b J_2+\frac{J_1^3}{3}-\frac{1}{2} J_2 J_1^2 \\ & \quad \;\;+J_1^2+\frac{1}{2} J_2^2 J_1-J_2 J_1+2 J_1+\frac{7 J_2^3}{6}+\frac{5 J_2^2}{2}+J_2+3
\end{aligned}$}
In this example we left some algebraic variables in the definitions to illustrate a feature of the package. For some properties, $\texttt{CIPro}$ is capable of computing quantities algebraically, without requiring integer inputs. However, this is not true of all properties one might desire, with a notable example being line bundle cohomology. The algorithm used to compute cohomology requires integer inputs.

\vspace{0.2cm}

Another application is to compute a derived vector bundle from an existent one. If, for example, one wanted to take the anti-symmetric power of the line bundle sum
\begin{eqnarray}
{\cal O}_X(1,1) \oplus {\cal O}_X(1,-1) \oplus{ \cal O}_X(-5,2) 
\end{eqnarray}
over the complete intersection Calabi-Yau
\begin{eqnarray}
X=\left[\begin{array}{c|ccc} \mathbb{P}^2&1&1&1\\ \mathbb{P}^4&2&2&1 \end{array} \right],
\end{eqnarray}
one could execute the following.
 \mybox{$\begin{aligned} 
   \text{In[]}&:=\text{CYandLBS}=<|\text{``DimPs''} \to \{2, 4\}, \text{``Conf''} \to \{\{1, 1, 1\},\{2, 2, 1\}\},\\& 
  \quad\quad\quad\quad\quad\quad\quad\quad \text{``LBS''} \to \{\{1,1,-5\},\{1, -1, 2\}\}|>;\\
  \text{In[]}&:=\text{CIMod}[\text{CYandLBS}, \text{``LBSWedge''},\text{``TensorPower''}\to 2] \\
  \text{Out[]}&:=<|\text{``DimPs''} \to \{2, 4\}, \text{``Conf''} \to \{\{1, 1, 1\},\{2, 2, 1\}\},\\& 
  \quad\quad\quad\quad\quad\quad\quad\quad \text{``LBS''} \to \{\{2, -4, -4\}, \{0, 3, 1\}\}|>;
\end{aligned}$}
Here the optional argument $\texttt{``TensorPower''}$ specifies what anti-symmetric power to take. As can be seen, this returns a new association, describing the desired wedge power.
\begin{eqnarray}
    \wedge^2 \left({\cal O}_X(1,1) \oplus {\cal O}_X(1,-1) \oplus{ \cal O}_X(-5,2)  \right) = {\cal O}_X(2,0) \oplus {\cal O}_X(-4,3) \oplus {\cal O}(-4,1)
\end{eqnarray}

\vspace{0.3cm}

The package can compute the following properties of line bundle sums: Direct sums, tensor products, symmetric and anti-symmetric powers, cohomology groups, maps between cohomology groups, cokernels, kernels and images of such maps, quotients, maps in the Koszul sequence, Chern classes, characters and their polynomials and index. Appendix \ref{app:modules} contains a complete list of the relevant keywords.

\section{Complete intersections with symmetries}
The package contains a number of tools to deal with finite (matrix) groups. The main purpose of implementing these is to facilitate defining CIs with finite automorphism groups which descend from the ambient  space. We start by reviewing the mathematical set-up.\\[2mm]
Recall that we are working in the multi-projective ambient space $\mathcal A = \PP^{n_1} \times \PP^{n_2} \times \cdots \times \PP^{n_m}$ with dimension $d=\sum_{i=1}^mn_i$ and homogeneous coordinates $x=(x^{(i)}_j)$, where $i=1,\ldots ,m$ and $j=0,\ldots , n_i$. We can think of this space as a toric variety: Starting with $\hat{\mathcal A}=\mathbb{C}^{n_1+1}\times\cdots\times\mathbb{C}^{n_m+1}=\mathbb{C}^{d+m}$ we obtain $\mathcal{A}$ by
\begin{equation}
\mathcal{A}=\frac{\mathbb{C}^{n_1+1}\setminus\{0\}\times\cdots\times\mathbb{C}^{n_m+1}\setminus\{0\}}{\mathcal{G}}
\end{equation}
that is, by taking out the `zero set' and dividing by the toric group $\mathcal{G}\cong(\mathbb{C}^*)^m$ with elements $(\lambda_1,\ldots ,\lambda_m)\in(\mathbb{C}^*)^m$ which act by coordinate re-scalings $x^{(i)}_j\mapsto \lambda_i x^{(i)}_j$ on each of the factors $\mathbb{C}^{n_i+1}$.
The package deals with finite groups  represented by (lists of) matrices $M$ with size $(d+m)\times (d+m)$ which are thought to act on the coordinates $x$. 

\subsection{Defining finite groups}
Finite groups are defined by a (finite) list $\{M_1,\ldots ,M_k\}$ of generators, each a matrix of size $(d+m)\times (d+m)$.

\subsubsection*{Examples}
As a simple example, for the ambient space $\mathcal{A}=\mathbb{P}^1$, we would like to set up the group generated by the matrices 
\begin{equation}
    M_1=\left(\begin{array}{rr}1&0\\0&-1\end{array}\right)\;,\qquad
    M_2=\left(\begin{array}{rr}0&1\\1&0\end{array}\right)\; .
\end{equation}
This is accomplished by executing
 \mybox{$\begin{aligned} \label{P1symm}
  \text{In[]}:=&\text{M1}=\{\{1,0\},\{0,-1\}\};\;\text{M2}=\{\{0,1\},\{1,0\}\};\\
  &\text{P1sym}=<|\text{"DimPs}\to\{1\},\text{"Conf"}\to\{\{\}\},\text{"Gen"}\to\{\text{M1},\text{M2}\}|>;
\end{aligned}$}
Groups can be defined `by hand' as just demonstrated but there is also a built-in data set of freely-acting symmetries of CICYs, as classified in Ref.~\cite{Braun:2010vc}. This dataset is called $\texttt{"Cicy3Sym"}$ and information about its structure can be obtained in the usual way, by executing
 \mybox{$\begin{aligned} 
  \text{In[]}:=&\text{Data}[\text{"Cicy3Sym"}]  
\end{aligned}$}
For concreteness, let us extract the generators for one of the bicubic symmetries by using the following command.
 \mybox{$\begin{aligned} \label{Z3Z3gen}
  \text{In[]}:=&\text{Z3Z3gen}=\text{Data}[\text{"Cicy3Sym"},\#[\text{"Conf"}]==\{\{3\},\{3\}\}\&][[2]][\text{"Gen"}];\\
  &\text{Z3Z3gen}//\text{MM}\\
  \text{Out[]}=&
\{\{
\left(
\begin{array}{cccccc}
 1 & 0 & 0 & 0 & 0 & 0 \\
 0 & e^{-\frac{2 i \pi }{3}} & 0 & 0 &
   0 & 0 \\
 0 & 0 & e^{\frac{2 i \pi }{3}} & 0 & 0
   & 0 \\
 0 & 0 & 0 & 1 & 0 & 0 \\
 0 & 0 & 0 & 0 & e^{-\frac{2 i \pi
   }{3}} & 0 \\
 0 & 0 & 0 & 0 & 0 & e^{\frac{2 i \pi
   }{3}} \\
\end{array}
\right),(1)\},
\{
\left(
\begin{array}{cccccc}
 0 & 0 & 1 & 0 & 0 & 0 \\
 1 & 0 & 0 & 0 & 0 & 0 \\
 0 & 1 & 0 & 0 & 0 & 0 \\
 0 & 0 & 0 & 0 & 0 & 1 \\
 0 & 0 & 0 & 1 & 0 & 0 \\
 0 & 0 & 0 & 0 & 1 & 0 \\
\end{array}
\right),(1)\}\}
\end{aligned}$}
The second argument of $\texttt{Data}$ can be a logical function to extract all data entries on which this function evaluates as $\texttt{True}$. In the above example, we have asked for all symmetries of the bicubic and have then extracted the generators of the second symmetry found in this way. The result is in fact a list of pairs of matrices, with the first entries the matrices acting on the coordinates $x$ as discussed above (for the bicubic ambient space $\mathcal{A}=\mathbb{P}^2\times\mathbb{P}^2$ this means $6\times 6$ matrices) and the second matrices acting on the defining sections of the CI. For the bicubic with only one defining section this is a $1\times 1$ matrix which happens to be trivial. 

\subsection{Calculating group properties}
The action of our finite group on the coordinates $x$ can be interpreted in two ways. If we think of $x$ as coordinates on $\hat{\mathcal A}$ then the matrices give rise to a linear group $G$ which acts on $\hat{\mathcal A}$. On the other hand, if $x$ represents homogeneous coordinates on $\mathcal{A}$ then the matrices describe a projective group $G_{\rm proj}$ which acts on $\mathcal{A}$. The group $G$ is known as the (multi-) Schur cover of $G_{\rm proj}$ and the two groups are related by
\begin{equation}
    G_{\rm proj}=\frac{G}{G\cap\mathcal{G}}\; .
\end{equation}
The package is capable of computing basic properties of the groups $G$, $G_{\rm proj}$ as well as of the sub-group $G\cap\mathcal{G}$ of $G$.

\subsubsection*{Examples}
Let us start with the simple example of the symmetry on $\mathcal{A}=\mathbb{P}^1$ defined in~\eqref{P1symm} and calculate properties of the linear group $G$ defined in this way.
 \mybox{$\begin{aligned} 
  \text{In[]}:=&\text{CIProp}[\text{P1sym},\{\text{"Order"},\text{"Abelian"}\}]//\text{MM}\\
  \text{Out[]}=&<|\text{"DimPs"}\to\{1\},\text{"Conf"}\to\{\{\}\},\text{"Gen"}\to \{
  \left(
\begin{array}{cc}
 1 & 0 \\
 0 & -1 \\
\end{array}
\right),
\left(
\begin{array}{cc}
 0 & 1 \\
 1 & 0 \\
\end{array}
\right)
\},\\
&\text{"Group"}\to \{
\{\left(
\begin{array}{cc}
 1 & 0 \\
 0 & -1 \\
\end{array}
\right)\},
\{\left(
\begin{array}{cc}
 0 & 1 \\
 1 & 0 \\
\end{array}
\right)\},
\{\left(
\begin{array}{cc}
 0 & -1 \\
 1 & 0 \\
\end{array}
\right)\},
\{\left(
\begin{array}{cc}
 1 & 0 \\
 0 & 1 \\
\end{array}
\right)\},\\
&\{\left(
\begin{array}{cc}
 -1 & 0 \\
 0 & 1 \\
\end{array}
\right)\},
\{\left(
\begin{array}{cc}
 -1 & 0 \\
 0 & -1 \\
\end{array}
\right)\},
\{\left(
\begin{array}{cc}
 0 & -1 \\
 -1 & 0 \\
\end{array}
\right)\},
\{\left(
\begin{array}{cc}
 0 & 1 \\
 -1 & 0 \\
\end{array}
\right)\}\},\\
&\text{"Order"}\to 8,
\text{"Abelian"}\to \text{False}\}
\end{aligned}$}
Evidently, the two generators listed with the keyword $\texttt{"Gen"}$ generate a non-Abelian group $G$ of order $8$, as indicated by the values of the keywords $\texttt{"Abelian"}$ and $\texttt{"Order"}$. The full group represented by the eight matrices listed under the keyword $\texttt{"Group"}$ is computed automatically, as one of the dependencies of $\texttt{"Order"}$. To find the properties of the projective group $G_{\rm proj}$ for this example we can carry out the following.
 \mybox{$\begin{aligned} 
  \text{In[]}:=&\text{P1symproj}=\text{CIProp}[\text{P1sym},\{\text{"ProjOrder","ProjAbelian"}\}];\\
  &\{\text{P1symproj["ProjOrder"]},
  \text{P1symproj["ProjAbelian"]}\}\\
  \text{Out[]}=&\{4,\text{True}\}
\end{aligned}$}
This means $G_{\rm proj}$ is an Abelian group of order $4$. In fact, closer inspection of the matrices shows that $G_{\rm proj}\cong\mathbb{Z}_2\times\mathbb{Z}_2$ and, as we have seen above, its Schur cover $G$ is non-Abelian and of order~$8$.\\[2mm]
As another example, let us use the generators in \eqref{Z3Z3gen} to define a symmetry on the bicubic CY.
 \mybox{$\begin{aligned} \label{bcZ3Z3}
  \text{In[]}:=&\text{bcs}=<|\text{"DimPs"}\to \{2,2\},\text{"Conf"}\to \{\{3\},\{3\}\},\text{"Gen"}\to \text{Z3Z3gen}|>;\\
  &\text{props}=\{\text{"Order", "Abelian", "ProjOrder", "ProjAbelian"}\};\\
  &\text{bcs} = \text{CIProp[bcs}, \text{props}];\; \text{Map[bcs[}\#]\&,\text{props}]\\
  \text{Out[]}=&\{27,\text{False},9,\text{True}\}
\end{aligned}$}
Hence, the Schur cover $G$ is a non-Abelian symmetry of order $27$ while its projective counterpart $G_{\rm proj}$ is Abelian and of order $9$. In fact, inspection of the matrices shows that $G_{\rm proj}\cong\mathbb{Z}_3\times\mathbb{Z}_3$.

\subsection{CIs with finite automorphism groups}
How does all this relate to CIs? Recall that the CI $X\subset\mathcal{A}$ is defined by the common zero locus of sections $s_a$, where $a=1,\ldots ,r$. Generalising the above set-up, group elements $g$ are actually represented by a pair, $g=(M,P)$ of matrices, with size $(d+m)\times (d+m)$ for $M$ and size $r\times r$ for $P$. If the matrix $P$ is omitted, it is assumed to be the unit matrix. Such pairs can act on the sections $s_a$ by
\begin{equation}
    g(s)_a(x)={P_a}^bs_b(M^{-1}x)\; .
\end{equation}
The matrix $P$ represents an automorphism of the bundle $\mathcal{N}$ and, in practice, is often a permutation matrix which permutes sections with the same degrees. If the sections are invariant, so $g(s)=s$ for all $g\in G$, then the complete intersection $X=\{x\in\mathcal{A}\,|\, s_1(x)=\cdots =s_r(x)=0\}$ has an automorphism group $G_{\rm proj}$.

\subsubsection*{Example}
Let us compute the most general defining polynomial for bicubic CYs with the symmetry $G_{\rm proj}\cong\mathbb{Z}_3\times\mathbb{Z}_3$, as defined in \eqref{bcZ3Z3}.
 \mybox{$\begin{aligned} 
  \text{In[]}:=&\text{CIProp[bcs},\{\text{"InvMon","DefEqs"}\},\text{"CoeffType"}\to\text{a}][\text{"DefEqs"}]\\
  \text{Out[]}=&
  \left\{a_{12} x_{1,0} x_{1,1} x_{1,2}
   x_{2,0} x_{2,1} x_{2,2} \right.\\
   &+a_4
   \left(x_{2,0} x_{2,1} x_{2,2}
   x_{1,0}^3+x_{1,1}^3 x_{2,0} x_{2,1}
   x_{2,2}+x_{1,2}^3 x_{2,0} x_{2,1}
   x_{2,2}\right)\\
   &+a_{10} \left(x_{1,1}
   x_{1,2}^2 x_{2,1} x_{2,0}^2+x_{1,0}
   x_{1,1}^2 x_{2,2}^2
   x_{2,0}+x_{1,0}^2 x_{1,2} x_{2,1}^2
   x_{2,2}\right)\\
   &+a_9 \left(x_{1,0}
   x_{1,1}^2 x_{2,1}
   x_{2,0}^2+x_{1,0}^2 x_{1,2}
   x_{2,2}^2 x_{2,0}+x_{1,1} x_{1,2}^2
   x_{2,1}^2 x_{2,2}\right)\\
   &+a_8
   \left(x_{1,0}^2 x_{1,2} x_{2,1}
   x_{2,0}^2+x_{1,1} x_{1,2}^2
   x_{2,2}^2 x_{2,0}+x_{1,0} x_{1,1}^2
   x_{2,1}^2 x_{2,2}\right)\\
   &+a_7
   \left(x_{1,1}^2 x_{1,2} x_{2,2}
   x_{2,0}^2+x_{1,0} x_{1,2}^2
   x_{2,1}^2 x_{2,0}+x_{1,0}^2 x_{1,1}
   x_{2,1} x_{2,2}^2\right)\\
   &+a_6
   \left(x_{1,0} x_{1,2}^2 x_{2,2}
   x_{2,0}^2+x_{1,0}^2 x_{1,1}
   x_{2,1}^2 x_{2,0}+x_{1,1}^2 x_{1,2}
   x_{2,1} x_{2,2}^2\right)\\
   &+a_5
   \left(x_{1,0}^2 x_{1,1} x_{2,2}
   x_{2,0}^2+x_{1,1}^2 x_{1,2}
   x_{2,1}^2 x_{2,0}+x_{1,0} x_{1,2}^2
   x_{2,1} x_{2,2}^2\right)\\
   &+a_3
   \left(x_{1,1}^3 x_{2,0}^3+x_{1,2}^3
   x_{2,1}^3+x_{1,0}^3
   x_{2,2}^3\right)+a_2 \left(x_{1,2}^3
   x_{2,0}^3+x_{1,0}^3
   x_{2,1}^3+x_{1,1}^3
   x_{2,2}^3\right)\\
   &+a_{11}
   \left(x_{1,0} x_{1,1} x_{1,2}
   x_{2,0}^3+x_{1,0} x_{1,1} x_{1,2}
   x_{2,1}^3+x_{1,0} x_{1,1} x_{1,2}
   x_{2,2}^3\right)\\
   &\left.+a_1 \left(x_{1,0}^3
   x_{2,0}^3+x_{1,1}^3
   x_{2,1}^3+x_{1,2}^3
   x_{2,2}^3\right)\right\}
\end{aligned}$}
The output is the most general polynomial defining a bicubic with automorphism group $G_{\rm proj}\cong\mathbb{Z}_3\times\mathbb{Z}_3$, where $a_i$ are arbitrary complex coefficients. That these coefficients are symbolic has been specified by the $\texttt{"CoeffType"}$ option. Omitting this options leads to the default which is random integer coefficients. To obtain this result for the defining equations, referred to by the keyword $\texttt{"DefEqs"}$, the invariant monomial combinations had to be computed first. This has been done by including the keyword $\texttt{"InvMon"}$ in the above example. Otherwise, in the absence of an $\texttt{"InvMon"}$ entry in the association, generic defining equations are generated. 

\section*{Acknowledgements}
A.L. is supported by the STFC consolidated grant ST/X000761/1. L.A and J.G. are supported, in part, by NSF grant PHY-2310588. The work of A.C. is
supported by the Royal Society grant DHF/R1/231142. The work of S.-J.L. is supported by the Yonsei University Research Fund 2026-22-0183. The authors are grateful to Albrecht Klemm for allowing us to incorporate his code into the present package and to Kit Fraser-Taliente for contributing some code to handle finite groups. The authors also thank their students and collaborators over the years who have used earlier versions of this package and identified issues.
\vskip 1cm
\appendix
\addcontentsline{toc}{section}{Appendix}
{\bf\Large Appendix}
\section{Overview of package structure} \label{app:struct}
Information about CIs and related objects is stored in Mathematica associations, which are data structures of the form \texttt{<|keyword1$\to$value1, keyword2$\to$value2,$\ldots$ |>}. A list of keywords can be found in Appendix~\ref{app:modules}. 
Introductory package help is obtained by executing \texttt{CIPro[]} or \texttt{CIPro["Long"]} and the present paper can be loaded by executing \texttt{CIPro["Manual"]}.
The package contains three main modules to compute with CI associations, namely the modules \texttt{CIProp}, \texttt{CIMod} and \texttt{Data}.\\[2mm]
The module \texttt{CIProp} is called with a CI association and a list of keywords and it computes the properties associated to those keywords. It returns a CI assocation with all the information of the input CI plus the computed properties (and any dependent properties which had to be computed first) which are appended in the  standard form \texttt{keyword}$\to$\texttt{value}.\\[2mm]
The module \texttt{CIMod} is also called with a CI association and a keyword and it modifies the CI as specified by the keyword. The output is a new association which describes the modified CI or, in some case, another data structure.\\[2mm]
The module \texttt{Data} loads in CI data supplied with the package. It is called with the name of the data set and a data selector which can be an integer, an integer range or a logical expression. It returns the part of the data set specified by the selector. A list of available data sets can be found in Appendix~\ref{app:data}.\\[2mm]
{\texttt{CIPro}} contains a number of auxiliary modules which can be listed by executing \texttt{?CISystemModules}. The most useful of those is the module \texttt{MM} which converts matrices inside expressions into matrix form. This command can, for example, be applied to CI associations and leads to a more easily readable form of these associations. Global package options can be displayed by executing $\texttt{Options[CIPro]}$.

\section{CIProp, CIMod and  keywords}\label{app:modules}
In this appendix, we describe the two main package commands \texttt{CIProp} and \texttt{CIMod} and their associated keywords.

\subsection{CIProp}
As discussed above, the module \texttt{CIProp} computes properties of CI associations. It can be called with different arguments. Executing \texttt{CIProp[]} without any arguments returns help for the command and  \texttt{CIProp["Long"]} returns a longer, more comprehensive version of this help which includes information on the mathematical background. Further, \texttt{CIProp["Options"]} and \texttt{CIProp["Keys"]} returns a list of options and keywords for the command, respectively. Help for any particular keyword \texttt{key} can be obtained by executing \texttt{CIProp[key]} or \texttt{CIProp[key,"Long"]}. To compute properties of an association one should execute\\[2mm]
\texttt{CIProp[association,listofkeys,<options>]}\\[2mm]
where \texttt{association} is a CI association, \texttt{listofkeys} is a keyword or a list of keywords and \texttt{options} are options (provided in the usual form \texttt{nameofoption$\to$value}). The command returns an association which contains all the information of the input associations plus the additional computed properties, added in the form \texttt{key$\to$value}, where \texttt{key} is any of the keys provided in the \texttt{listofkeys} argument. The command takes dependencies into account automatically. If the calculation of one of the properties specified in \texttt{listofkeys} depends on the prior calculation of another property this will be determined automatically (and the result will be added to the output association) even if the key for this property is not contained in \texttt{listofkeys}. A list of property keys and their meaning is provided in Tables~\ref{tab:CIkeys}, \ref{tab:Symmkeys}, \ref{tab:CISymmkeys} and \ref{tab:lbskeys}.
\begin{table}[!h]
 \centering
 \begin{tabular}{|l|l|l|}\hline
 key&meaning&possible options\\\hline\hline  
 \texttt{"AmbCoords"}&homogenous coordinates for $\mathcal{A}$&\texttt{"CoordName"$\to$symbol}\\\hline
 \texttt{"AmbDim"}&dimension of ambient space $\mathcal{A}$&\\\hline
 \texttt{"C"}&CI Chern class&\\\hline
 \texttt{"Ch"}&CI Chern character&\\\hline
 \texttt{"ChPoly"}&CI Chern character as a polynomial&\\\hline
 \texttt{"CIDim"}&CI dimension&\\\hline
 \texttt{"Conf"}&CI configuration matrix&\\\hline
 \texttt{"CPoly"}&CI Chern class as a polynomial&\\\hline
  \texttt{"CY"}&CI is a CY, \texttt{True} or \texttt{False}&\\\hline
 \texttt{"DefEqs"}&CI defining polynomials&\texttt{"CoeffType"$\to$symbol}\\
 &&\texttt{"CoeffRange"$\to$\{min,max\}}\\
 &&\texttt{"Eqs"$\to$listofpolys}\\\hline
 \texttt{"DimPs"}&dimension of projective factors&\\\hline
 \texttt{"Euler"}&CI Euler number&\\\hline
  \texttt{"GVInv"}&GV invariants for CICYs&\texttt{"MaxDeg"$\to$integer}\\\hline
  \texttt{"HilbertSeries"}&CI Hilbert series&\texttt{"HilbertFormat"$\to$format}\\\hline  
 \texttt{"ISec"}&CI intersection numbers&\\\hline
 \texttt{"ISecForm"}&CI intersection form&\\\hline
 \texttt{"K"}&CI canonical bundle&\\\hline
 \texttt{"MonList"}&CI defining monomials&\\\hline
 \texttt{"Mu"}&CI integrating form&\\\hline
 \texttt{"Singular"}&CI is singular, \texttt{True} or \texttt{False}&\\\hline
 \texttt{"SingularLocus"}&def.~eqs.~for CI nodal variety&\\\hline
 \texttt{"Td"}&CI Todd class&\\\hline
  \texttt{"TdPoly"}&CI Todd class as a polynomial&\\\hline
\end{tabular}
\caption{A list of keywords related to CI properties for the command \texttt{CIProp}.}
    \label{tab:CIkeys}
\end{table}
\begin{table}[!h]
\centering
\begin{tabular}{|l|l|}\hline
  key&meaning\\\hline\hline 
  \texttt{"Abelian"}&is matrix group $G$ Abelian, \texttt{True} or \texttt{False}\\\hline
  \texttt{"Gen"}&generator of matrix group\\\hline
 \texttt{"GenOrder"}&a list of oders of generators\\\hline
 \texttt{"Group"}&the group $G$ as a list of matrices\\\hline
 \texttt{"Order"}&the order of $G$\\\hline
 \texttt{"PermGroupGen"}&permutation group associated to $G$\\\hline
 \texttt{"ProjAbelian"}&is projective group $G_{\rm proj}$ Abelian, \texttt{True} or \texttt{False}\\\hline
 \texttt{"ProjGroup"}&the group $G_{\rm proj}$ as a list of matrices\\\hline
 \texttt{"ProjOrder"}&order of $G_{\rm proj}$ \\\hline
 \texttt{"ProjTrivialGroup"}&projectively trivial sub-group $G\cap\mathcal{G}$ of $G$ \\\hline
 \texttt{"ProjTrivialOrder"}&order of  $G\cap\mathcal{G}$\\\hline
\end{tabular}
\caption{List of keywords related to symmetry properties for the command \texttt{CIProp}.}
\label{tab:Symmkeys}
\end{table}
\begin{table}[!h]
\centering
\begin{tabular}{|l|l|}\hline
   key&meaning\\\hline\hline 
   \texttt{"ConfInvQ"}&configuration invariant under symmetry, \texttt{True} or \texttt{False}\\\hline
   \texttt{"InvMon"}&$G$-invariant defining monomial combinations for CI\\\hline
   \texttt{"InvQ"}&defining eqs.~invariant under $G$, \texttt{True} or \texttt{False}\\\hline
\end{tabular}
\caption{List of keywords related to CI and symmetry properties for the command \texttt{CIProp}.}
\label{tab:CISymmkeys}
\end{table}
\begin{table}[!h]
\centering
\begin{tabular}{|l|l|l|}\hline
   key&meaning&options\\\hline\hline 
   \texttt{"AmbCohDim"}&ambient space line bundle cohomology dimension&\\\hline
   \texttt{"Basis"}&monomial basis for ambient line bundle cohomology&\\\hline
   \texttt{"ChLBS"}&Chern character of LBS sum&\\\hline
   \texttt{"ChLBSPoly"}&Chern character of LBS as polynomial&\\\hline
   \texttt{"CLBS"}&Chern class of LBS&\\\hline
   \texttt{"CLBSPoly"}&Chern class of LBS as polynomial&\\\hline
   \texttt{"CohInfo"}&Information about LBS cohomology&\texttt{"CohInfo"}$\to$list\\\hline
   \texttt{"CohList"}&list of $q_i$ in $H^{q_i}(L_i)$ for LBS $\bigoplus_iL_i$&\texttt{"CohQ"}$\to$list\\\hline
   \texttt{"Coset"}&coset matrices for ambient LBS cohomology&\\\hline
   \texttt{"IndLBS"}&index of a LBS&\\\hline
   \texttt{"LBS"}&LBS as a matrix, each column a line bundle&\\\hline
   \texttt{"LBSCohDim"}&LBS cohomology dimensions&\\\hline
   \texttt{"LBSRank"}&rank of a LBS&\\\hline
\end{tabular}
\caption{List of keywords related to CI and line bundle sum properties for the command \texttt{CIProp}.}
\label{tab:lbskeys}
\end{table}
\subsection{CIMod}
The module \texttt{CIMod} computes new CI associations from given ones. As with \texttt{CIProp}, it can be called with different arguments. Executing \texttt{CIMod[]} without an argument or \texttt{CIMod["Long"]} displays a standard or long form of help for the module. As before, \texttt{CIMod["Options"]} and \texttt{CIMod["Keys"]} returns information about the available options and keywords, respectively. Help on any particular keyword \texttt{key} is obtained by executing \texttt{CIMod[key]} or \texttt{CIMod[key,"Long"]}. The standard way to use the module is\\[2mm]
\texttt{CIMod[associationlst,opkey,<options>]}\\[2mm]
where \texttt{associationlst} is an association or a list of associations (depending on the operation performed) and \texttt{opkey} is a string specifying the operation. The output is a new CI association or, in some case, a different data structure. A list of keywords and associated options for \texttt{CIMod} can be found in Table~\ref{tab:CIMod}.
\begin{table}[!h]
\centering
\begin{tabular}{|l|l|l|}\hline
   key&meaning&options\\\hline\hline 
   \texttt{"CINormal"}&normal form for CI configuration matrix&\\\hline
   \texttt{"CIEquiv"}&are two CI configurations equivalent?&\\\hline
   \texttt{"LBSSum"}&sum of LBSs&\\\hline
   \texttt{"LBSTensor"}&tensor product of LBSs&\texttt{"TensorPower"}$\to$int\\\hline
   \texttt{"LBSWedge"}&antisymmetric product of LBSs&\texttt{"TensorPower"}$\to$int\\\hline
   \texttt{"MapLBS"}&random map between two LBSs&\\\hline
   \texttt{"MapSpaceLBS"}&Ker, Im, Coker for LBS cohomology map&\texttt{"Space"}$\to$string\\
   &&\texttt{"Map"}$\to$matrix\\
   &&\texttt{"CoeffType"}$\to$string\\
   &&\texttt{"CoeffRange"}$\to$range\\\hline
   \texttt{"QuotientLBS"}&quotient of LBS cohomologies&\\\hline
   \texttt{"KoszulLBS"}&maps in Koszul complex for LBSs&\texttt{"Ekpq"}$\to\{k,p,q\}$\\\hline
   \texttt{"CohLBS"}&line bundle cohomology&\texttt{"CohFormat"}$\to$string\\
   &&\texttt{"CohProt"}$\to$prottype\\
   &&\texttt{"CohSum"}$\to$boolean\\\hline
 \end{tabular}
\caption{List of keywords for the command \texttt{CIMod}.}
\label{tab:CIMod}
\end{table}

\section{Data}\label{app:data}
The module \texttt{Data} loads in CI data provided with the package. Executing \texttt{Data[]} with no argument returns help on how to use the module and a list of available data sets which can also be found in Table~\ref{tab:data}. The format for all data sets is as a list of CI associations. To obtain help for any of these data sets execute \texttt{Data[dataname]}.
The standard way to use the module is\\[2mm]
\texttt{Data[dataname,cond,<listofkeys>]}\\[2mm]
where \texttt{dataname} is the name of the data set, and \texttt{cond} specifies which parts of the data set to extract. In the simplest case, \texttt{cond} is an integer $n$ to extract the 
$n^{\rm th}$ entry from the data set. To extract a range of entries, \texttt{cond} should be of the form $\{n_{\rm min},n_{\rm max}\}$ and the entire data set is loaded by setting it to \texttt{All}. The length of the data set is obtained by setting \texttt{cond} to \texttt{"Length"}. Finally, \texttt{cond} can be a Boolean function and all data entries for which this function evaluates to \texttt{True} are returned. The argument \texttt{listofkeys} is a key or a list of keys and, if provided, only the properties associated to those keys are returned for all CI associations which satisfy the condition \texttt{cond}. 
\begin{table}[!h]
\centering
\begin{tabular}{|l|l|}\hline
  name of data set&description of content\\\hline\hline 
  \texttt{"Cicy3"}&list of 7890 CICY 3-folds from Ref.~\cite{Candelas:1987kf}\\\hline
 \texttt{"Cicy3Favour"}&list of equivalent favourable CICY 3-folds from Ref.~\cite{Anderson:2017aux}\\\hline
 \texttt{"Cicy3Sym"}&a list of freely-acting CICY 3-fold symmetries from Ref.~\cite{Braun:2010vc}\\
 & Hodge numbers for smooth quotients taken from Refs.~\cite{Constantin:2016xlj, Candelas:2016fdy} \\\hline
 \texttt{"Cicy3FavourSym"}&freely acting symmetries for favourable Cicys from Ref.~\cite{Gray:2021kax}\\\hline
 \texttt{"Cicy4"}&list of $921497$ CICY 4-folds from Refs.~\cite{Gray:2013mja,Gray:2014kda}\\\hline
 \texttt{"delPezzo2"}&CI realisations of del Pezzo surfaces\\\hline
 \texttt{"LBSStandardModels"}&list of promising $SU(5)$  line bundle models from Refs.~\cite{Anderson:2013xka,Constantin:2018xkj}\\\hline
  \texttt{"LBSStandardModels202"}&list of heterotic line bundle standard models from Ref.~\cite{Anderson:2011ns}\\\hline
 
\end{tabular}
\caption{List of data sets available with the package.}
\label{tab:data}
\end{table}

\section{Computing line bundle cohomology}  \label{app:lbc}
In this appendix, we sketch the algorithm for computing line bundle cohomology used in the \texttt{CIPro} package. This algorithm is based on the Bott--Borel--Weil description of line bundle cohomology (see, for example, Ref.~\cite{hubsch}) on projective spaces, the Koszul sequence and an interation of the associated spectral sequence. (For further mathematical background see, for example, Refs.~\cite{schenck,cox,gh,h}.)\\[2mm]
The problem we would like to address is as follows. Consider the ambient space
\begin{equation}\label{Adef1}
\mathcal A = \PP^{n_1} \times \PP^{n_2} \times \cdots \times \PP^{n_m}\; ,
\end{equation}
a line bundle $\mathcal{L}={\cal O}_{\mathcal{A}}({\bf k})$ on $\mathcal{A}$, as in Eq.~\eqref{lbamb}, a co-dimension $r$ CI variety $X\subset\mathcal{A}$ with dimension $d$, associated bundle $\mathcal{N}=\bigoplus_{a=1}^r{\cal O}_{\mathcal{A}}({\bf q}_a)$, as defined in Eq.~\eqref{Xdef}, and the line bundle $L=\mathcal{L}|_X$ obtained from the ambient line bundle $\mathcal{L}$ by restriction to $X$. The goal is to determine a suitable representation (and the dimension) of line bundle valued cohomology groups $H^q(X,L)$ of $L$.\\[2mm]
We provide here an overview of the algorithm (for more detailed accounts see \cite{hubsch,Anderson:2008ex}). A key ingredient is the Koszul sequence
\begin{equation}\label{koszul}
    0\longrightarrow \wedge^r\mathcal{N}^*\otimes\mathcal{L}\stackrel{\wedge^rs}{\longrightarrow }\cdots \stackrel{\wedge^3s}{\longrightarrow}\wedge^2\mathcal{N}^*\otimes\mathcal{L}\stackrel{\wedge^2s}{\longrightarrow}\mathcal{N}^*\otimes\mathcal{L}\stackrel{s}{\longrightarrow}\mathcal{L}\stackrel{\rho}{\longrightarrow} L\longrightarrow 0
\end{equation}
where $\rho$ is the restriction from $\mathcal{L}$ to $L$ and the other maps are given in terms of the sections $s=(s_1,\ldots ,s_r)$ whose common zero locus defines the CI $X$, as in Eq.~\eqref{Xdef}. This sequence, and its associated spectral sequence, facilitate computing the desired cohomology groups $H^q(X,L)$ in terms of ambient space cohomology groups $E_0^{p,q}=H^q(\mathcal{A},\wedge^p\mathcal{N}^*\otimes\mathcal{L})$ and sub-spaces and quotients within these. The cohomologies of ambient space line bundle sums are represented following the Bott--Borel--Weil formalism and sub-spaces and quotients within by using standard linear algebra methods.\\[2mm]
In order to determine the line bundle cohomologies $H^q(X,L)$ we have to compute the spectral sequence $(E_k^{p,a})$ whose initial entry is given by the spaces $E^{p,q}_0$ defined above. This is done by iteration, $E_k^{p,q}\rightarrow E_{k+1}^{p,q}$, where $E_{k+1}^{p,q}$ is given by  the cohomology of the previous entry $E_k^{p,q}$, equipped with maps $f_k^{p,q}:E_k^{p,q}\rightarrow E_k^{p-k,q-k+1}$. These  maps are the restrictions of maps
\begin{equation}\label{F_def}
 F_k^{p,q}:H^q({\cal A},\wedge^p\mathcal{N}^*\otimes {\cal L})\rightarrow H^{q-k+1}({\cal A},\wedge^{p-k}\mathcal{N}^*\otimes {\cal L})\; ,
\end{equation} 
which, in turn, are induced from the maps in the Koszul sequence~\eqref{koszul}. Starting with $E_0^{p,q}$ the spectral sequence is iterated in this way until, after at most $r$ steps, it stabilises to $E_\infty^{p,q}$. The desired line bundle cohomology groups are then obtained from
\begin{equation}
     H^q(X,L)\cong \bigoplus_{i\geq 0}E_\infty^{i,q+i}\; .
\end{equation}
In simple cases, the maps in \eqref{F_def} are effectively those of the Koszul sequence (e.g. maps between $H^{q}({\cal A},\wedge^{p}\mathcal{N}^*\otimes L)\to H^{q}({\cal A},\wedge^{p-1}\mathcal{N}^*\otimes L)$ for which $k=1$ above). In some cases however, the induced maps are more subtle and require more involved relationships between the elements~$E_k^{p,k}$.\\[2mm]
Perhaps the least straightforward aspect of the algorithm outline above is the construction of the maps $F_k^{p,q}$ in Eq.~\eqref{F_def}. For the most part, these maps are obtained from the defining sections $(s_1,\ldots  ,s_r)$ of the CI, by taking suitable anti-symmetric tensor powers according to the power $\Lambda^p \mathcal{N}^*$ of $\mathcal{N}^*$ involved. However, matters are somewhat more complicated for maps related to co-boundary maps and we would like to present an example for how to deal with such a case.\\[2mm]
Consider the CICY 3-fold defined by the configuration matrix
\begin{equation}
 X=\left[\begin{array}{l|lll} \mathbb{P}^1&1&0&1\\\mathbb{P}^1&0&1&1\\ \mathbb{P}^4&1&3&1\end{array}\right]\; ,
\end{equation} 
and the line bundle $L={\cal O}(0,0,5)$. For the computation of this cohomology, a non-trivial map is induced from $F_3^{3,2}:H^2({\cal A},\wedge^3\mathcal{N}^*\otimes{\cal L})\rightarrow H^{0}({\cal A}, {\cal L})$ (which includes co-boundary maps from a C\v ech cohomology viewpoint) or, explicitly, $F_3^{3,2}: H^2({\cal A}, {\cal O}(-2,-2,5)) \to H^0({\cal A},{\cal O}(0,0,5))$.

Here the Bott--Borel--Weil formalism indicates that the morphism from the 1-dimensional source space to the target cohomology group can be written in explicit polynomial form, but it is not simply the defining equations themselves, but a more intricate double determinant of their coefficients. We write down the three defining polynomials ($s_1,s_2,s_3)$ as
\begin{equation}
(s_1,s_2,s_3)=(s^{\alpha i}x_\alpha z_i,s^{aijk}y_az_iz_jz_k,s^{\beta b l}x_\beta y_{b} z_{l})\; ,
\end{equation}
where $x_\alpha$, $y_a$, $z_i$ are the homogenous coordinates of the three ambient projective space factors $\mathbb{P}^1,\mathbb{P}^1$, and $\mathbb{P}^4$ respectively. With this notation, the map takes the form
\begin{equation}
F_3^{3,2}=\epsilon_{\alpha \beta}\epsilon_{ab}s^{\alpha i}s^{ajkl}s^{\beta b m}z_{i}z_{j}z_{k}z_{l}z_m
\end{equation}
where $\epsilon_{\alpha\beta}$ and $\epsilon_{ab}$ are both the totally antisymmetric tensor in 2 dimensions. For further details see Refs.~\cite{hubsch,Anderson:2008ex}.

\end{document}